\documentclass{elsart}

\usepackage{graphicx,amssymb}

\begin{document}

\begin{frontmatter}

\title{The Cryogenic Target for the G$^0$ Experiment at Jefferson Lab} 

\author[aau]{S. D. Covrig\corauthref{cor}},
\corauth[cor]{Corresponding author.}
\ead{silviu@krl.caltech.edu}
\author[aaz]{E. J. Beise},
\author[aau]{R. Carr}, 
\author[aau]{K. K. Gustafsson}, 
\author[aau]{L. Hannelius}, 
\author[aaz]{M.-C. Herda},
\author[aau]{C. E. Jones},
\author[aaz]{J. Liu},
\author[aau]{R. D. McKeown},
\author[aac]{R. Neveling\thanksref{rnow}},
\thanks[rnow]{iThemba Labs, PO Box 722, Somerset West, South Africa}
\author[avv]{A. W. Rauf}, 
\author[aav]{G. Smith}

\address[aau]{California Institute of Technology, Pasadena, CA, USA}
\address[aaz]{Physics Department, University of Maryland, College Park, MD \ 20472, USA}
\address[aac]{Loomis Laboratory of Physics, University of Illinois, Urbana, IL \ 61801, USA}
\address[aav]{Thomas Jefferson National Accelerator Facility, Newport News, VA \ 23606, USA}
\address[avv] {TRIUMF, 4004 Wesbrook Mall, Vancouver, BC V6T 2A3, Canada}

\date{\today} 

\begin{abstract}

A cryogenic horizontal single loop target has been designed, built, tested and 
operated for the G$^0$ experiment in Hall C at Jefferson Lab. The target cell is 
20 cm long, the loop volume is 6.5 l and the target operates with the cryogenic
pump fully immersed in the fluid. The target has been designed to operate at
30 Hz rotational pump speed with either liquid hydrogen or liquid deuterium. 
The high power heat exchanger
is able to remove 1000 W of heat from the liquid hydrogen, while the nominal electron
beam with  current of 40 $\mu$A and energy of 3 GeV deposits about 320 W of heat 
into the liquid. 
The increase in the systematic uncertainty due to the liquid hydrogen target is negligible 
on  the scale of a parity violation experiment. The global normalized yield 
reduction for 40 $\mu$A beam is about 1.5 $\%$ and
the target density fluctuations contribute less than 238 ppm (parts per million) to the
total asymmetry width, typically about 1200 ppm, in a Q$^2$ bin.

\end{abstract}

\begin{keyword}
Liquid hydrogen target, unpolarized targets \sep Density variations \sep Parity violation 
\PACS  29.25.-t \sep 25.30.Bf
\end{keyword}
\end{frontmatter}



\section{Introduction}
    
In the G$^0$ experiment \cite{g0prop} in Hall C at Jefferson Lab, the parity-violating 
(PV) asymmetry in elastic electron scattering from hydrogen and quasi-elastic electron 
scattering from deuterium is measured in the Q$^2$ range from 0.1 
to 1 (GeV/c)$^2$ in both forward and backward angle modes by scattering a longitudinally 
polarized electron beam on unpolarized liquid targets. By measuring three independent 
asymmetries, one at forward angles on liquid hydrogen and two at backward angles, one 
on liquid hydrogen and one on liquid deuterium, a complete separation and mapping of 
the strange vector form factors of the nucleon (G$_M^s$, G$_E^s$) and the isovector 
axial form factor (G$_A^e(T=1)$) in three Q$^2$ bins over the Q$^2$ range from 0.1 to 
1 (GeV/c)$^2$ can be performed for the first time. In the forward angle mode the particle 
detected is the proton, while in the backward angle mode the particle detected is the 
electron.

The measured PV asymmetries are in the range of 1 to 30$\times$10$^{-6}$, with 
the intended contribution from systematic uncertainties of less than 5$\%$ of 
the measured asymmetry.
In order to make the physics program possible the target system must accommodate large
power depositions into the liquid with the minimum possible systematic uncertainties
due to global yield reduction and density fluctuations on the time scale of 30 Hz of 
the asymmetry measurement.

The target is inside the vacuum enclosure of a superconducting magnet
system (SMS) \cite{g0prop}, which poses several challenges. The target had
to be designed to fit inside a cylindrical volume with a 60.96 cm diameter 
(this being the space constraint imposed by the magnet's inner core). The 
target system must be reliable during several months continuous operation,
and should have no interferences with the toroidal magnetic field. 

In order to accomplish this ambitious experimental program a recirculating closed 
loop high power cryogenic target system has been designed, 
built, thoroughly tested, installed and operated. In order to meet the space constraints
the cryogenic loop is placed horizontally inside the SMS and is fed horizontally
with gases. The electron beam interacts with the target fluid in the target cell
and the cryogenic fluid circulates in a high speed flow in the horizontal
loop. The nominal beam spot on target is a square of side 2 mm. The target
provides a cylindrical clearance for the beam with a diameter of roughly 12.7 mm.
The cryogenic loop has 2 degrees of freedom in a plane perpendicular to the beam
line and no degree of freedom along the beam line.

The remainder of this paper is organized as follows: section 2 provides a detailed 
description of the target system, section 3 contains the tests carried out on the 
target with and without beam, and section 4 is a summary of operational
experience with beam on target and a summary of measured target parameters.

\section{The target system}

The major components of the target system are: the cryogenic loop, the service
system and the controls. 

\subsection{Cryogenic loop}

\begin{figure}[htb]
\centering
\rotatebox{90}
{\includegraphics[scale=0.6]{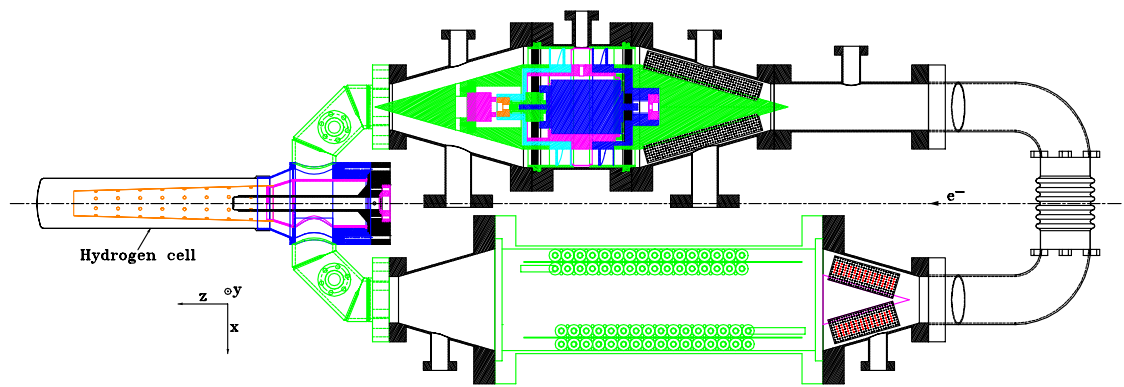}}
\caption{The cryogenic loop. The ports along the loop
are not in the mounted position.}
\label{fig:loop}
\end{figure}

The cryogenic loop, as shown in fig. \ref{fig:loop}, can be divided into 
several units: the cryogenic pump housing, the heat exchanger housing, the target 
manifold and the bellows. The target manifold is the only unit made of Al.
The shell of the rest of the loop is made of Type 304 stainless steel. 
During normal running the loop is placed horizontally inside the superconducting 
magnet, in the
same plane as the beam line. The orientation of the loop with respect to the 
direction of the beam line is with the target cell placed downstream. To ensure
safe beam operation on target, the upstream part of the loop goes below the
horizontal plane and contains a bellows that connects the two sides of the loop 
below the beam plane (see fig. \ref{fig:smodule}). Except for the upstream 
bellows all loop parts are rigidly interconnected. The bellows reduces the 
mechanical stress due to the rigid connections along the loop. Once the loop 
is put together, the bellows is clamped in place and the whole cryogenic loop 
becomes a rigid structure.

\subsubsection{The target manifold}

\begin{figure}[htb]
\centering
\rotatebox{90}
{\includegraphics[scale=0.6]{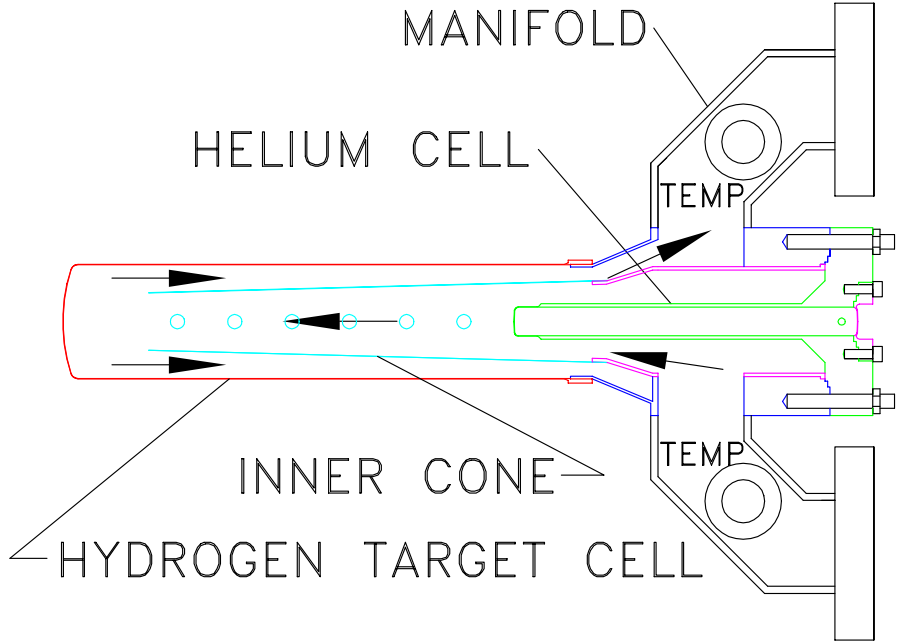}}
\caption{The target manifold.}
\label{fig:manifold}
\end{figure}

The target manifold, shown in fig. \ref{fig:manifold}, is completely made of Al-6061 T6 
and connects with the loop
through transition stainless-Al conflat flanges. All loop units are connected 
through conflat flanges with Cu gaskets, rated for high vacuum (10$^{-11}$ Torr) well
beyond that needed in the experiment (the vacuum is nominally around 10$^{-6}$ Torr 
during normal operations).

The target manifold houses two cells, a primary hydrogen cell and a secondary
helium cell. The primary cell is a thin shell cylinder, about 23 cm long, 
soldered to an Al joint downstream of the manifold. The cell was machined
in one piece from Al-6061 T6 with an outer shell thickness of 0.178 mm and an
internal diameter of 5 cm. 
The helium cell is 16 cm long and is placed upstream of the primary
hydrogen cell. The helium cell is a cylindrical tube machined in one piece on
a flange that connects to the manifold through an indium seal. Its inner
diameter is 12.7 mm, which defines the clearance volume for the electron beam.
The tube of the helium cell has a nipple soldered at the downstream end.
The nipple is machined in one piece and has a thin window of 0.228 mm thickness. 
At the upstream end, the helium cell is protected against vacuum by an Al thin 
window, 0.178 mm thick, machined on an Al flange that connects to the helium cell
flange through an indium seal. The helium cell is operated at the same pressure
as the hydrogen cell. There are three thin Al windows in beam.
The distance between the exit window of the helium cell and the exit window of 
the hydrogen cell is 20 cm and defines the liquid hydrogen target in beam.
Nominal running conditions with liquid hydrogen are 1.7 atm and 19 K, 3 K below
its boiling point. With these conditions, the target thickness
in beam is 1.44 g/cm$^2$, yielding luminosities in excess of 2$\cdot$10$^{38}$ 
cm$^{-2}$s$^{-1}$.

There are two particular reasons for having the helium cell. The exit window 
of the helium cell has the same shape and radius of curvature as the exit window 
of the hydrogen cell. The thin exit window of the hydrogen cell has a spherically 
convex shape with a radius of curvature of 7.6 cm. Hence, systematic effects 
caused by parallel beam drifts on target are reduced to first order. Another 
feature of the helium cell design is that it provides azimuthal symmetry for 
scattered particles, originating in the hydrogen target, in the range of the
angular acceptance of the SMS.

\subsubsection{The heat exchanger}\label{sec:hx}

One of the cryogenic loop's legs houses a double coil counterflow high power
heat exchanger. The coils are made of finned Cu tubing on the hydrogen side to 
increase the area for heat exchange. The heat exchanger uses cold helium gas 
as coolant agent, from Jefferson Lab's End Station Refrigerator (ESR), with
nominal parameters of 15 K and 12 atm. The heat removal is on average 50 W
for a coolant flow of 1 g/s.
The inner space of the heat exchanger's coils is filled with an Al flow
diverter (not shown in fig. \ref{fig:loop}) that forces the target fluid to go 
through the fins. The area for heat exchange on the hydrogen side is 9500.73 
cm$^2$, and on the coolant side is 1110 cm$^2$. The effective diameter 
for flow in the hydrogen circuit is 30.4 cm$^2$ and 0.338 cm$^2$ on the coolant 
side.

\subsubsection{Cryogenic pump}

The other leg of the cryogenic loop houses the cryogenic pump and the
high power heater (HPH). The pump is a vane-axial design with two impellers 
in series. The impellers have three blades each and are rigidly attached 
on a motor shaft that is immersed into the cryogenic fluid. The volume 
dislodged by the pump in one revolution, corrected for the volume of 
the blades, is 0.198 l. The cryogenic motor that is  presently in use is a 
Barber-Nichols Inc. custom DC brushless sensorless motor, 
driven by a sensorless controller. The motor is rated for liquid 
hydrogen and has a 12.7 mm diameter shaft. Downstream from the pump there is a 
conical Al-made flow diverter inside which there is a tachometer. The 
tachometer is completely separate from the pump system. It is made of
a Cu coil rigidly attached to the Al flow diverter. The downstream
pump impeller has a permanent small field magnet dipole rigidly attached to 
it. When the impeller rotates the varying magnetic flux through the coil,
caused by the magnet, induces a varying electric field with the same 
frequency  as the rotational frequency of the impeller and offers 
a measure of the motor's rotation that is independent of the 
motor's controller. The tachometer signal is read back by a
digital multimeter enabled in frequency mode (Agilent model 34401A). 

Upstream of the pump there is a second conical flow diverter made
of Al that houses the HPH. The purpose of the HPH is to regulate the
heat load on the target in a feedback loop during normal operations.
The HPH is made of three independent coils wrapped on a G10 support, 
connected in parallel at the back of the power supply. In this configuration, 
in case one of the coils becomes inoperable, it doesn't incapacitate 
the whole heater. A reconfiguration of the remaining
coils at the back of the power supply will enable its continual 
operation. The coils are made of Ni-Cr alloy ribbon and each has a 
resistance of about 3.5 $\Omega$, giving a total 
resistance of 1.15 $\Omega$. The HPH power supply, PowerTen 3300P/4025,
is operated remotely with two low-voltage dc control signals 
(0-10 V = 0-100 $\%$) that are set by analog output channels from
a VME board.

There are flow diverters inside the cryogenic loop, all made
of Al-6061. The purpose of the flow diverters is to guide the fluid
smoothly around the loop and assure a high Reynolds number in
the target cell and heat exchanger, where turbulence facilitates
heat transfer or heat removal and mixing. In addition to the flow 
diverters along the loop, inside the target cell there is a 
conical tube made of Al-1100, of 0.0762 mm thickness (see
fig. \ref{fig:manifold}). The tube acts
as a flow diverter and guides the liquid down the center of the target 
cell serving two purposes. The conical geometry increases the flow speed 
as it traverses the cell parallel to the beam direction, increasing the
turbulence and mixing of the fluid. The conical tube
has holes on the side that run along the surface. The holes allow the 
fluid to leave the interaction region before reaching the end of the 
cell, helping to remove heated fluid from the beam path sooner and 
reducing the intrinsic
interaction time between the beam and the fluid. The holes also
serve a mechanical purpose, they prevent the conical tube tearing
by relieving the static pressure on the inside of the conical tube.
The holes sit in the shadows of the superconducting magnet coils, and 
thus do not influence the scattered particles within the experimental 
acceptance.

With the exception of the cryogenic motor rotor, all materials used in the 
fabrication and assembly of the target cryogenic loop are either
low magnetic susceptibility or non-magnetic materials. The rotor
of the cryogenic motor is made of strong permanent rare-earth magnets
with fields at the surface in excess of 1 kGauss, which decay quickly
with radial distance. The motor axis sits about 10 cm off the central 
beam axis in the horizontal plane, with the center of the motor about 
46 cm from the center of the target cell. Measurements of motor's 
magnetic field done with Hall probes found that the magnetic field half 
way between the bearings and 5 cm from the rotor's axis of revolution is 
less than 7 Gauss. At 10 cm the field is less than 1 Gauss. The field 
along the rotor's axis of revolution is 1.5 Gauss at one bearing and 
less than 1 Gauss at the other bearing. The magnetic field integral
along a particle's trajectory changes the nominal field integral
of the SMS by about 0.01 ppm, which is negligible for this measurement.

\subsubsection{Instrumentation}

Along the cryogenic loop there are eight electrical feedthroughs with 
mini-conflat flanges used for instrumenting six temperature sensors 
directly into the fluid, the tachometer, the heaters and the 
cryogenic motor. 
The six temperature sensors used on the loop are LakeShore Cernox 
CX-1070-AA. 
An additional set of temperature sensors of the same kind
were instrumented in the helium coolant lines with two sensors 
right across the heat exchanger to monitor the coolant temperature 
to and from the target. The temperature sensors are read by Oxford 
ITC-502 temperature controllers. The temperature sensors are distributed 
along the cryogenic loop in pairs, two across the pump and HPH, two 
across the target manifold and two across the heat exchanger. 
The electrical lines from the loop go through a second set of feedthroughs
on the service module to the G$^0$ electronics racks in the hall. 
The cables are about 42 m long and made of Kapton insulated Cu. 
As an extra precaution the wires under vacuum have been shielded with 
4-channels ceramic beads. The beads are radiation and fire resistant. 

\subsection{The service system}

The service system for the target is split in two major components,
the mechanical service module, that supports and moves the target,
and the gas handling system.

\subsubsection{The gas handling system}

\begin{figure}[htb]
\centering
\rotatebox{0}
{\includegraphics[scale=0.70]{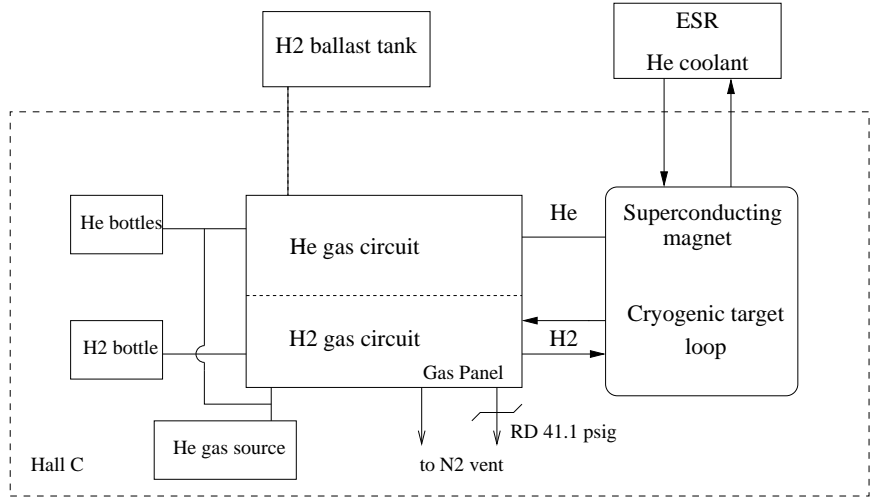}}
\caption{The gas handling system.}
\label{fig:gaspanel}
\end{figure}

Fig. \ref{fig:gaspanel} shows a schematic of the gas handling system 
(the dashed line is the Hall C boundary), consisting of gas service
lines, a gas panel and a ballast tank.

The external coolant supply lines are vacuum insulated coaxial pipes. 
The coolant comes from the 15 K - 12 atm cold helium supply at the ESR. 
The flow needed to extract heat loads of up to 500 W on the target is 
less than 14 g/s. The coolant flow to the target's heat exchanger is 
controlled by a needle valve, which is manually operated.

The gas panel sits in experimental Hall C and its primary purpose is 
to supply the cryogenic loop with gases. It supplies both cells in the 
cryogenic loop from dual
supplies (the hydrogen cell can be supplied with gas from a bottle in
the hall or from a ballast tank outside the hall; the helium cell can
be supplied with helium from bottles in the hall or from a 4 atm source 
of high purity helium). In addition to supplying gases to the two
cells in the cryogenic loop, the gas panel connects the hydrogen cell to 
adequate relief paths for excessive pressure, either to the ballast 
tank or to the hall's standard dry nitrogen vent. The panel's functions 
are possible through a combination of manual, pneumatic, solenoid, check 
and relief valves. 
On the gas panel there are 2 pneumatic valves controlled by a vacuum 
interlock box. The interlock box prevents the operation of these valves 
unless there is sufficient vacuum in the magnet's vessel. One pneumatic valve
connects the gas panel with a pumping station, the other one connects the
gas panel with the ballast tank during normal running. The interlock
box prevents pumping on the cryogenic loop unless adequate vacuum
is established in the magnet's vessel, hence protecting the thin
hydrogen cell against implosion. In case of a vacuum breach in the 
magnet vessel, the interlock box also closes the connection between the 
ballast tank and the cryogenic loop, reducing the inventory of hydrogen 
(deuterium) that might come in contact with air. 
There are three relief paths in parallel on the gas panel  
two instrumented with relief valves that open at 25 and 30 psig 
respectively, and one with a rupture disk that breaks at 41.1 psig.

The gas panel is instrumented with pressure gauges for direct reading
of pressures in the system and with pressure transducers for remote
monitoring of pressures. The transducers, made by Omega, are of three varieties, 
one absolute with range from 0 to 100 psi, another one absolute with a range from
0 to 3000 psi, and one differential with a range from -15 to 15 psid.
The differential transducers are used to read the pressure across the
exit window of the helium cell and the pump head.

All the components on the gas panel are either explosion proof or rated
for hydrogen. To prevent accidental remote operation all the components 
of the gas handling system can only be monitored from a computer.

In the normal mode of operation, before a run 
starts, the cryogenic loop and the helium cell are pumped and purged several 
times (following \cite{userman}) with helium before the running fluids are made
available to the target. Before cool-down starts, the cryogenic loop is
put in connection with the ballast tank and will stay connected to it
at all times during a run, unless a vacuum incident triggers the vacuum
interlock box to close the connection. The ballast tank acts as a big
pressure reservoir and mitigates pressure excursions in the target 
during normal running. The ballast tank is also a storage reservoir 
for target fluid (hydrogen or deuterium), it has a volume of 2500 gallons.
As a safety precaution, some critical valves in the gas delivery system 
for the target are locked and only target experts have access to them.

\subsubsection{The service module}

The target's service module (SM) has been manufactured by Thermionics NW.
The basic functions of the SM are to support the cryogenic loop, to provide 
motion control along two axes perpendicular to the beam axis and to
\begin{figure}[htb]
\centering
\rotatebox{0}
{\includegraphics[scale=0.7]{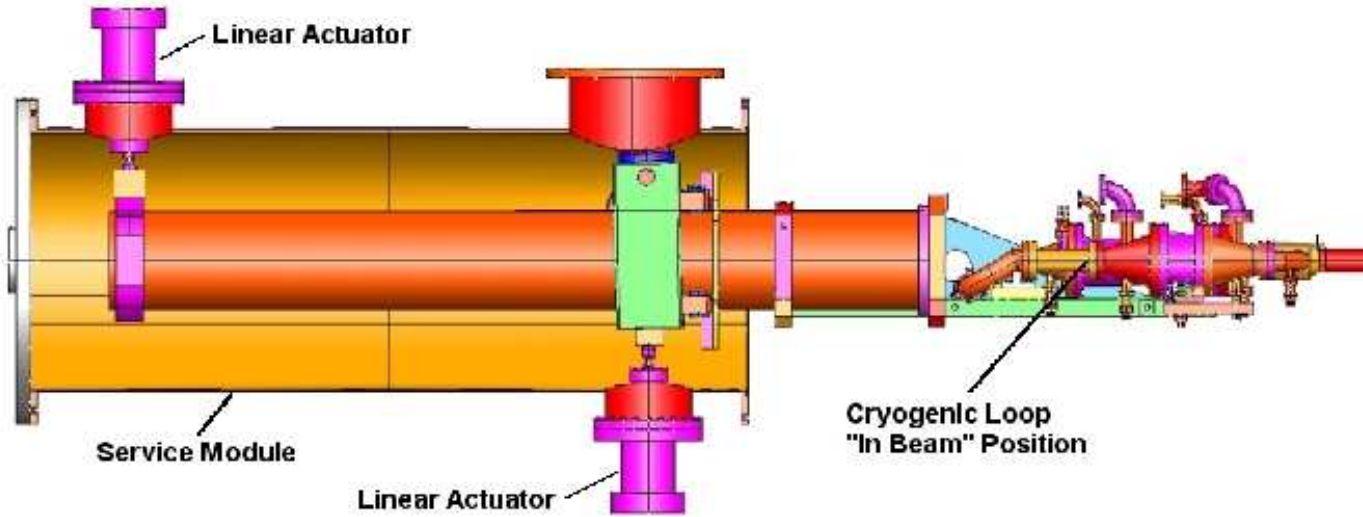}}
\caption{The service module. The target is shown in the \it{``in beam''} position.}
\label{fig:smodule}
\end{figure}
provide an interface for gas lines and electrical lines from vacuum to the
outside world. A schematic view of the SM can be seen in fig. 
\ref{fig:smodule}, along with the functionality of the associated motion system.
The cryogenic loop is rigidly supported inside the SMS on a cantilevered 
platform from the SM. The cantilever is moved by four linear actuators 
through ball joints. The actuators are placed in pairs in two vertical planes at different
locations along the beam axis. With these degrees of freedom the target can
be placed in position, aligned to the magnet-beam axis to better than 1 mm in
both x and y directions (the nominal reference system has the z axis along the
beam line with positive sense down the line, the y axis is vertical with positive
sense up, and the x axis is horizontal in-plane with positive sense to beam right).
The actual location precision of the motion system along a single axis is 0.05 mm, 
and for two axes combined it should position the target to better than 0.1 mm.

The SM is located on the beamline upstream of the SMS. It interfaces with the 
SMS through a 60.96 cm diameter flange and with the beamline through a 15.24 
cm gate valve. The SM cannot be separated from the SMS without breaking the 
vacuum, but it can be separated from the beamline with the gate valve. The 
target cell can be moved about 7.6 cm vertically and about $\pm$2.5 cm 
horizontally. In the vertical direction the target has an out of beam position 
with 2.5 cm clearance for beam. To protect the target from hitting the inner 
bore of the SMS, under any circumstances, the motion system is instrumented 
with limit switches arranged on a ring around the beam axis. Once a switch
hits a wall the motion in that direction is blocked and only retraction is 
possible.

\subsection{Target controls}

The target controls system \cite{controls} is a set of electronics, computers 
and software used to monitor, operate and control the target. The main functions 
of the controls system are monitoring target parameters, warn the operator of
critical conditions, log target data at regular intervals, provide
selected target data to the online data acquisition system, control
the target heaters to ensure target stability and move the target along two
axes perpendicular to the beam axis (the target system has no degree of
freedom along the beam line). The controls for this target have been described 
in detail in [3]. The monitoring and controlling electronics sit in
two electronics racks in the hall, shielded from direct radiation from the 
target. The signals are centralized in the hall into an Input-Output Controller
(IOC) that is located in a VME crate in a rack in the hall. The VME processor
is a mv2700 power PC. The analog input and output to the IOC  is done through
Greenspring ADIO modules. The IOC is also able to process data from serial ports.
Information from the IOC is shipped through the local Ethernet to a monitoring
computer located in the Hall C Counting House. This computer is the monitoring
and control station for target parameters during normal running and is operated by a 
trained target operator. All the target monitoring and controlling electronics
are on Uninterruptible Power Supplies (UPS) to ensure safe monitoring and
controlling of the target system in case of a power failure.
The target data are part of the larger EPICS database at Jefferson Lab.
The operation of the target is carried out through a combination of software, 
and hardware. The cryogenic pump speed is controlled only through a linear 
potentiometer, and not via computer, in order to avoid interruption of the
pump's rotation due to IOC and/or computer crashes.

The HPH is primarily controlled through a Proportional-Integral-Differential
(PID) feedback loop with a temperature sensor in the cryogenic loop.
In our implementation of the PID loop the feedback tracks the beam
current incident on the target and subtracts the deposited beam power 
from the total power on the target, setting the rest on the heater and 
keeping a temperature sensor constant. This helps stabilize the target 
quicker with less than 0.2 K temperature excursions during beam trips. 
A parallel complete manual control of the HPH is used for situations when
the IOC feedback loop is nonoperational (due to network failures or IOC crashes).
All other target parameters, besides pump speed, gas panel and sometimes
the HPH, are controlled through the IOC. All target parameters are monitored
through the IOC on the target computer screen. The target controls system is 
based on EPICS \footnote{http://www.aps.anl.gov/epics} and MEDM 
\footnote{http://www.aps.anl.gov/epics/extensions/medm}.

\section{Target tests and performance}

There are two principles at the base of our testing program, safety and performance.
The target had been very throughly tested in order to be certified as safe, 
since it operates explosive gases (hydrogen and deuterium).
The target design performance needed to be demonstrated before installing
it in the experimental hall. 
The preliminary target checks and tests included mechanical integrity tests 
(pressure tests, leak checks etc.) and performance tests with cold helium 
gas in the cryogenic loop. 

\subsection{Pre-beam tests}

The purpose of the pre-beam tests were to assess the target overall dynamical
performance, safe operation and reliability. These tests were split into 
three different sets: cold helium tests, liquid neon test and liquid 
hydrogen tests. 

\subsubsection{Cold helium performance}\label{sec:coldhe}
 
In the Test Lab at Jefferson Lab the target was tested for the first 
time in cryogenic conditions by using helium coolant (4 K / 3 atm at 
maximum 5 g/s) delivered by the Cryogenic 
Test Facility (CTF) refrigerator and helium gas as target fluid in the cryogenic loop. 
The cold helium tests were meant primarily to assess the cryogenic pump performance
and the operation of the target's instrumentation. During the period of these
cold helium tests the target has been filled with the coldest cryogen to date, 
7 K / 20 psig helium gas (about a quarter of the pump load expected with liquid 
hydrogen). The pump efficiency, eq. \ref{eqn:eff}, and mass flow, eq. \ref{eqn:qdot},
can be determined by measuring the 
fluid  temperature increase as it goes across a known heat source \cite{beise},
which in this case is the HPH. In this approach helium is better than hydrogen 
as it yields larger temperature differences for the same heat load than hydrogen 
and the pump efficiency is measured with better precision. However, the cold helium
is in gaseous state and is compressible as opposed to hydrogen which would be liquid
and incompressible (see fig. \ref{fig:pump_perf}a for a comparison of the measured pump 
efficiency between helium and hydrogen). The heat exchanger was also tested for 
heat transfer performance. The coolant used in the Test Lab
differs significantly from the one in the Hall (CTF delivers cold helium at
4 K / 3 atm, while the ESR delivers cold helium at 15 K / 12 atm). In
these conditions the heat exchanger removed in excess of 500 W of
heat loads from the target's fluid.

The cryogenic pump geometry has been designed to insure high speed fluid
flow at a nominal rotational frequency of 30 Hz. Ample fluid flow is
necessary in order to reduce target density related effects and mitigate
systematic errors in the asymmetry measurements. To sustain the fluid
flow needed it was determined, after a set of initial tests, that a high
torque custom motor was necessary for powering the cryogenic pump.
The Barber \& Nichols motor proved to have the desired performance and 
is the motor presently powering our cryogenic pump. Although it runs with 
no problems at 30 Hz in liquid hydrogen, it saturates in current at about 
42.7 Hz and it may be just marginal for reaching 30 Hz in liquid deuterium.

\subsubsection{The neon test}

The target cells have been pressure tested to 85 psid. The estimate for 
maximum cell pressure in case of a liquid hydrogen catastrophic boil-off 
(the most serious scenario was considered
to be the sudden loss of vacuum in the magnet vessel to room temperature air)
was 29 psia with the designed venting lines. To test the design of the relief
system for the cryogenic target a simulated loss of vacuum was performed 
using neon as target fluid and dry room temperature nitrogen as $''air''$. 
Neon has several qualities which make it suitable for such a test. It is 
inert, it has a liquefaction point close to hydrogen (28.98 K for neon and 
22.21 K for hydrogen at 1.7 atm - the pressure for normal
running conditions in the target), and is much denser than liquid hydrogen.
The fact that the liquefaction points are close for the two fluids makes
temperature differences from liquid state to room temperature to be similar
for the two fluids, an important consideration when simulating loss of
vacuum to room temperature air. Because liquid neon is 17 times denser than 
liquid hydrogen, the expected pressure peak in a vacuum loss was 38 psia,
a 30\% increase compared with hydrogen. The length in time expected
for the neon peak pressure to last is more than double the peak pressure 
time length expected for a catastrophic boiling of liquid hydrogen. 
For this test the target was mounted on the SMS and it successfully passed the 
neon test, \emph{in situ}.
The cryogenic motor was able to push liquid neon in the cryogenic loop 
saturating in frequency at 12.2 Hz, the saturation in torque, eq. \ref{eqn:torque}, 
was determined to be about 23 oz-in (the same torque limit reached with liquid
hydrogen at 42.7 Hz).

\subsubsection{Liquid hydrogen performance}

For a week in early June 2002, the target underwent basically the same set of 
tests it did with cold helium, this time with liquid hydrogen.

As outlined in Section \ref{sec:coldhe}, the pump flow can be determined by measuring 
the temperature difference of the fluid as it passes across a known heat source:

\begin{equation}
\eta{\cdot}\dot{Q}_H = \dot{m}_{{LH}_2}\cdot c_P \cdot (T_{PO} - T_{PI})
\label{eqn:qdot}
\end{equation}

where $\dot{Q}_H$, in our case, is the power delivered by the high power heater 
power supply and $\eta{\cdot}\dot{Q}_H$ is the power deposited
by the HPH into the loop. The parameter $\eta$ accounts for
the power loss in the electrical supply cables. The logged parameter was 
the power delivered by the power supply. To get the power deposited into the
loop, $\eta$ was determined to be 0.75.

Data were taken at four different cryogenic pump rotational frequencies: 
10, 20, 30 and 40 Hz, and at six different heater power values: 550, 
450, 350, 250, 150 and 50 W for each frequency. The specific isobaric
heat and density for liquid hydrogen were corrected for temperature
dependence. Since the pressure stayed the same during these studies,
there were no pressure related corrections to the input hydrogen
parameters. At each power value the data included in the computation
of the mass rate were the closest to the thermodynamic equilibrium.
With these data the pump efficiency can be determined using the relation
\begin{equation}
\epsilon_P = \frac{\eta}{c_P(T){\cdot}{\rho(T)}{\cdot}V_s{\cdot}f}{\cdot}\frac{\dot{Q}_H}{T_{PO} - T_{PI}}
\label{eqn:eff}
\end{equation}
where $V_s$ is the theoretical volume displacement of the cryogenic
pump in one revolution. The cryogenic pump has two identical impellers, 
each equipped with three blades.
The estimate for $V_s$, taking into account the volume of the blades,
is 0.198 liters. The rotational frequency of the cryogenic motor shaft
is denoted by \emph{f}. The motor frequency was measured with the
tachometer. The temperature was measured with the temperature sensors
instrumented along the loop. The temperature across the HPH is denoted
by $T_{PO} - T_{PI}$, where indexes mean pump-out and pump-in,
respectively.

With the differential pressure across the cryogenic pump one can
determine the torque of the cryogenic motor at different frequencies.
The relation between torque and pressure is
\begin{equation}
\tau = \frac{V_s{\cdot}\Delta p}{2\pi}
\label{eqn:torque}
\end{equation}
and the mechanical power delivered by the cryogenic pump can be 
determined from the data with the relation
\begin{equation}
P = \epsilon_P{\cdot}V_s{\cdot}f{\cdot}\Delta p.
\end{equation}
For a counterflow heat exchanger, like the one used in G$^0$, the log 
mean temperature difference \cite{heattr} is given by
\begin{equation}
\Delta T_{LM} = \frac{T_{ho} - T_{ci} - (T_{hi} - T_{co})}{ln({\frac{T_{ho} - T_{ci}}{T_{hi} - T_{co}}})}
\label{eqn:delta_tlm}
\end{equation}
and the heat exchanger coefficient is given by
\begin{equation}
U = \frac{\dot{q}_{He}}{\Delta T_{LM}} = \frac{\dot{m}_{He}{\cdot}c_p^{He}{\cdot}(T_{co} - T_{ci})}{\Delta T_{LM}},
\label{eqn:hxcoeff}
\end{equation}
where T$_{hi}$ is the temperature of the liquid hydrogen going into the
heat exchanger and T$_{co}$ is the temperature of the helium
coolant coming out of the heat exchanger, for example.
During the heat exchanger studies the coolant mass flow
as read from the ESR flow-meter for the 15 K supply for 
Hall C never exceeded $\dot{m}_{He} = $10.5 g/s. As in the pump studies
$\Delta T_{LM}$ was determined for four cryogenic pump
frequencies and six setpoints for the high power heater.

To characterize the heat exchanger further its effectiveness
has been determined, which for a counter flow heat exchanger is
defined theoretically \cite{heattr} as
\begin{equation}
\epsilon_{HX}^{th} = \frac{1-e^{-N{\cdot}(1-R)}}{1-R{\cdot}e^{-N{\cdot}(1-R)}}
\end{equation}
where R is the heat capacity rate ratio, and N is the
number of heat transfer units, defined as
\begin{equation}
\nonumber R = \frac{{(\dot{m}{\cdot}c_P)}_{min}}{{(\dot{m}{\cdot}c_P)}_{max}},\quad  N = \frac{U}{{(\dot{m}{\cdot}c_P)}_{min}} 
\end{equation}
where \emph{min/max} refer to the helium coolant or hydrogen liquid,
$(\dot{m}{\cdot}c_P)_{min} = min({\dot{m}}_H{\cdot}c^H_P,{\dot{m}}_{He}{\cdot}c^{He}_P)$.
Physically it represents the ratio between the actual rate
of heat transfer in the heat exchanger and the maximum
allowable rate by the second law of thermodynamics. 
Experimentally the value of the effectiveness is given by
\begin{equation}
\epsilon_{HX}^{exp} = \frac{T_{co} - T_{ci}}{T_{hi} - T_{ci}}
\end{equation}

\subsubsection{Results and comments}

\begin{figure}[htb]
\centering
\rotatebox{-90}
{\includegraphics[scale=0.45]{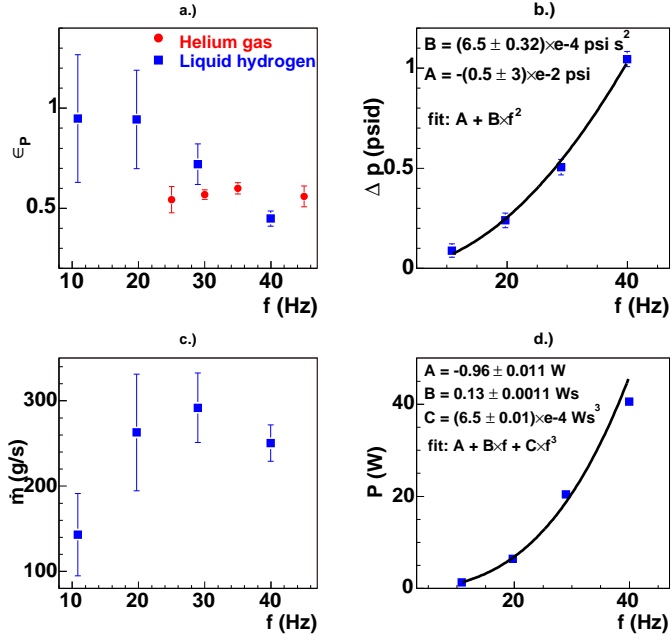}}
\caption{Cryogenic pump performance in liquid hydrogen.}
\label{fig:pump_perf}
\end{figure}

The cryogenic pump efficiency, fig. \ref{fig:pump_perf}a, and 
flow measurements, fig. \ref{fig:pump_perf}c, in the loop have 
large uncertainties due to the small temperature differences in 
hydrogen that can be induced across the HPH (on average about 
only four times the systematic calibration error of 25 mK). Data 
from the two lowest heater setpoints (50 and 150 W) have been 
excluded in averaging the pump efficiency and mass rate. With the 
present controller the cryogenic pump saturated in current at 42.7 
Hz. At the saturation torque, of 23 oz$\cdot$in, the cryogenic pump 
head was 1 psid, and the mechanical power delivered to the liquid 
was about 43 W. It can be inferred from the data on the saturation 
frequency in liquid neon and liquid hydrogen that the saturation 
frequency is inversely proportional with the square root of the 
density of the fluid. Based on this inference the predicted saturation 
frequency in liquid deuterium is about 29 Hz at the nominal running 
point in this liquid. The power delivered by the pump is fitted well 
by a polynomial of third degree in frequency, fig. \ref{fig:pump_perf}d. 
The pump head, fig. \ref{fig:pump_perf}b, is fitted well by a second 
degree polynomial in frequency, although the pump efficiency is 
frequency dependent in liquid hydrogen. The mass flow seems to be 
maximum at the nominal frequency of 30 Hz, with a pump efficiency 
of 0.7. The average flow speed in the liquid hydrogen cell was 
determined to be about 8 m/s. The Reynolds number for the liquid 
hydrogen flow determined in these conditions is in excess of 10$^6$. 
Since the transition between laminar flow and turbulent flow happens 
for flows with Reynolds numbers between 2000 and 10000, the liquid 
hydrogen flow in our target at 30 Hz is well into the turbulent flow 
region.
\begin{figure}[htb]
\centering
\rotatebox{0}
{\includegraphics[scale=0.45]{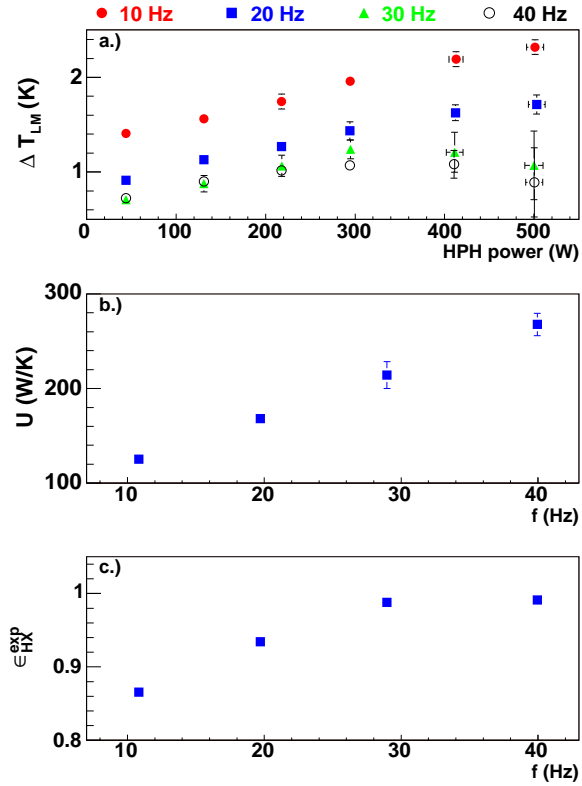}}
\caption{Heat exchanger performance in liquid hydrogen.}
\label{fig:hx_perf}
\end{figure}

The heat exchanger removed 450 W of heat from the loop with a coolant flow 
of 10.5 g/s from the 15 K helium coolant supply. The flow of the coolant 
has 30$\%$ uncertainty.
The heat transfer coefficient, U was computed with the semi-empirical 
formula
\begin{equation}
{\frac{1}{U}} = {{\frac{1}{(h_{He}{\cdot}A_{He})}} + {\frac{1}{(h_{LH_2}{\cdot}A_{LH_2})}}} 
\label{eqn:U}
\end{equation}
where A is the area for heat transfer on helium/liquid hydrogen side respectively 
and
\begin{equation}
h = 0.023{\cdot}\frac{C_P{\cdot}G^{0.8}{\cdot}{\eta}^{0.2}}{D_e^{0.2}{\cdot}{Pr}^{0.6}} 
\label{eqn:h}
\end{equation}
where G is the mass flow per unit area of the flow, $\eta$ (g/cm s) is the 
fluid's dynamic viscosity, D$_e$ is the wetted perimeter and Pr is the Prandtl 
number (dimensionless) for the fluid (Pr = $\eta{\cdot}$C$_P$/k, where k (W/cm K) 
is the thermal conductivity). For nominal running conditions at a cryogenic 
pump speed of 30 Hz  the heat transfer coefficient was estimated to be U = 
285 W/K for a heat exchanger with no leakage. At the same pump speed the data 
yielded a value for the heat transfer coefficient 
of U = 214 W/K, fig. \ref{fig:hx_perf}b. The discrepancy comes from the model 
computed value for U, which assumes that the temperature is constant across 
the fins, while in reality it is a gradient from the helium side to the tip of 
the fins, and the gradient varies along the heat exchanger. Also, 
originally the heat exchanger had nylon rope wrapped around the
fin tubing to force the liquid hydrogen flow to go around the
fins, but the rope was removed because of concerns that it may 
block the flow on the hydrogen side, and due to this the
heat exchanger has some leakage that depends on the flow speed.
The leakage has the effect of reducing the value of the heat transfer 
coefficient compared to that from eqs. (\ref{eqn:U}-\ref{eqn:h}).

The mean temperature difference across the heat exchanger,
$\Delta T_{LM}$ has the expected behavior in frequency, it
decreases with the increase in pump frequency, see fig. \ref{fig:hx_perf}a,
but it is not linear at high rotational frequencies of the cryogenic 
pump and high heat loads which is also related to the heat exchanger's
leakage. The effectiveness, see fig. \ref{fig:hx_perf}c,
is very close to the theoretical estimate
for a counter flow heat exchanger, for N$>3$ it should be above
0.95. In our heat exchanger the number of heat transfer units
is at least 3.55, and the measured effectiveness is above 0.98 
for pump frequencies higher than 30 Hz. 
The heat exchanger was tested for maximum heat transfer in Dec 2003 
and up to the maximum power on the HPH (1000 W) was able to remove 
the heat load from the liquid hydrogen, while keeping the target
at the nominal running point (19 K, 1.7 atm). The helium coolant
flow was about 80\% of the maximum deliverable by the ESR on the 
15 K supply, so in principle this type of heat exchanger is able to
remove more than 1000 W of heat load from liquid hydrogen.

After the liquid hydrogen was vaporized, when the temperature in the 
cryogenic loop was about 50 K, a cryogenic pump frequency scan was 
performed and the pump turned up to 73 Hz. The controller was turned 
off, and the shaft stopped turning in 23 seconds, as read on the 
tachometer readback, meaning that there is no significant friction 
impeding the rotation in cryogenic conditions.

\subsection{In-beam tests}

A set of tests with beam on target was carried out, scanning the location 
of the target in the beam to check the alignment, calibrating the HPH and 
checking the PID feedback loop with beam. The target location was scanned 
by stopping the beam, moving the target a known amount in each of the 
transverse directions to the beam direction, restoring the beam and recording 
the current from a background scintillator detector on a photomultiplier tube, 
located close to the target. For a beam size of 2 by 2 mm on the target, it was 
found that the target cell axis was centered on the beam axis to within 1 mm 
in both x and y, in agreement with the alignment specs for the experiment. 
From the HPH calibration it was determined that at full beam on target,
40 $\mu$A, the target consumes about 320 W of cooling power. Theoretical
estimates of target heating from a 40 $\mu$A, 3 GeV electron beam are
around 300 W (Al windows included). The difference of 20 W is cryogenic 
pump heating and losses to the environment. 

\subsubsection{Liquid hydrogen operation}

The target is normally operated with liquid hydrogen at 19 K, or 3 K below 
its boiling point at 1.7 atm. The cryogenic pump
rotated at 31 Hz with speed fluctuations of less than 0.1 Hz over months
long continuous running. Since the helicity signal used to compute the PV
asymmetry in this experiment changes on a 30 Hz basis
the pump was run at a slightly different frequency from 30 Hz in order to avoid
possible sources of systematics in the asymmetry measurements due to the
pump rotation. The amplitude of the hydrogen cell vibrations due to the 
pump's rotation at full speed (75 Hz) was measured and found to be below 
0.01 mm. Since these vibrations do not happen at 30 Hz it is not expected
that they correlate with the helicity signal. The electron beam's natural 
motion on target is about ten times bigger than the cell's vibrations. 
The contribution from the cell's vibrations to the beam position on target 
is then expected to be negligible. The stability of the cryogenic pump rotation
over time is a strong indication that the pump induced hydrogen cell 
vibrations are also stable in time. Target instabilities with beam trips are 
well mitigated by the HPH on PID. The temperature excursions seen when the 
beam trips or comes on target are less than 0.2 K and last about 20 s. The 
liquid hydrogen relative density change due to beam trips is 0.3$\%$. 
The relative normalized yield change in our detectors is about 1$\%$. 
The difference is due to the fact that the beam, when ramped up to nominal 
current, also drifts in space by as much as a few mm. To avoid this problem
a beam trip cut was implemented in the data acquisition that excludes about 25 s
of data after the beam comes on target. With relatively stable
beam on target (40$\pm$0.5 $\mu$A) the PID maintains the temperature
constant to within 0.02 K (relative target density change expected
from PID stabilization is 0.03\%, completely negligible for this
experiment). 

\subsubsection{Boiling studies}

Dedicated data were acquired during two separate engineering runs to study
the density fluctuations on the time scale of the asymmetry measurement.
The detection technique in G$^0$ is based on a time of flight spectrum.
Because of this G$^0$ does not use the standard CEBAF beam structure at 
499 MHz, but rather a modified version of it at 31 MHz. This makes it 
possible to separate the elastic events from inelastics in the time of 
flight spectra. However,
G$^0$ is a parity violation (PV) experiment that ultimately measures 
not a cross section but an asymmetry \cite{g0prop}. From the target 
point of view there are two effects that influence the systematics in this 
kind of measurement. Since the
target is a liquid it can have density fluctuations and global density
reduction. Density fluctuations are random changes in the liquid density 
driven mainly by either fluctuations in beam properties (like intensity, 
position etc.) or random effects due to the interactions between beam
particles and the target liquid (microscopic bubble formation, window
heating etc.). The mechanisms for density fluctuations are
complex and realistic cases are poorly understood quantitatively. Their 
effect translates directly into an enlargement of the measured asymmetry 
width. This enlargement in turn is a source of additional uncertainty in the
asymmetry measurement and it thus increases the asymmetry error bar. 
Density reduction is a global reduction in the density of the liquid
whenever there is beam on target. What it translates into is the reduction
of the measured normalized yield from the liquid (as the effective target
thickness is smaller) and in turn has the effect of reducing the statistics
accumulated in a given period of time. With experiments on liquid 
targets, neither of the above mentioned density effects is avoidable. 
Therefore, it is necessary to reduce them until they become negligible 
on the time scale of the helicity change, 30 Hz for G$^0$. 

The basic computing unit for the PV asymmetry in this experiment 
is a quartet. A quartet stands for four consecutive helicity
states, either having a structure like $+--+$ or $-++-$. The measured
asymmetry is formed between the sum of normalized yield with positive
helicity and the sum of normalized yield with negative helicity, in the
same quartet, divided by the sum of all four. The detector yields
have been normalized to beam charge.
The 40 $\mu$A G$^0$ electron beam is very focused 
in space, nominally about 200 microns in both x and y. Not only would this beam
boil a liquid target in its path but it may also damage the target system 
as it has power densities in the range of kW/mm$^2$. To mitigate this the beam is
rastered uniformly over a larger area. In this experiment the raster shape 
is a square, nominally with a side of 2 mm, or an area of 4 mm$^2$. Two
triangular signals are swept along x and y with frequencies f$_x$ = 24.96 kHz
and f$_y$ = 25.08 kHz with a linear speed of 1000 m/s to yield a square
pattern with a uniformity of 95\%. The goal of the  target density studies 
was to determine the extra width due to target density
fluctuations on the measured asymmetry width at the nominal running point
in the experiment, beam at 40 $\mu$A, 3 GeV, rastered at 2x2 mm$^2$ and the
cryogenic target pump rotating at 31 Hz. 

\subsubsection{Density fluctuations}\label{sec:density}

The target density fluctuations depend on target medium temperature,
cryogenic pump rotational speed and beam raster size, intensity and 
intrinsic spot size. Of all these dependencies measurements were made
to assess target density fluctuations versus raster size, pump speed and
target medium temperature. During each measurement only one parameter
at a time was varied.

During the first engineering run, the G$^0$ focal plane detector 
(FPD \cite{g0prop}) was segmented in 
two statistically independent detectors to check if the measured effect 
was concurrent. To separate deadtime corrections associated with time 
encoding electronics, which are rather complicated, from target density
fluctuations the electronics were used in the scaler mode. In this mode
the electronics counts detector hits with a deadtime of about 30 ns.
The total asymmetry width, when the whole detector was summed up, was
320 ppm in nominal beam and pump running conditions.
The density fluctuations were studied at constant beam current by changing 
the raster and this was repeated at different pump speeds. In the second 
engineering run data were taken with the same conditions as in the first 
engineering run (except for a different target cell). A newly instrumented 
detector set, the luminosity detectors or lumis, was also used in the second 
run. There were 8 lumi detectors placed in two sets of 4 at different locations 
along the beam line sensitive to very forward scattering, mostly M$\o$ller 
electrons and elastic electrons from e-e and e-p scattering in the target
(from the beam line lumis 1-4 at 2$^\circ$ to the target, clocked at 
45$^\circ$ in the x-y plane, and lumis 5-8 at 1.2$^\circ$ and along x-y axes, 
one up, one down, one left, one right).

\begin{figure}[htb]
\centering
\rotatebox{-90}
{\includegraphics[scale=0.32]{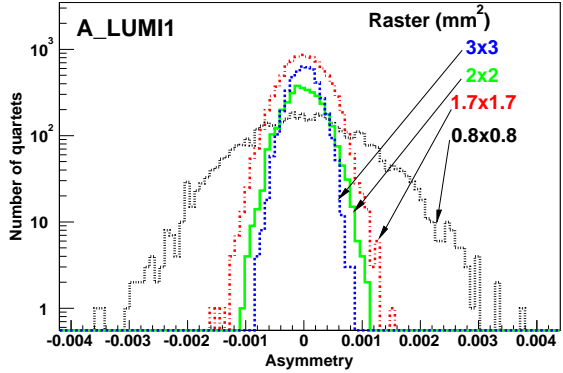}}
\caption{Lumi1 asymmetry width at 40 $\mu$A beam and different raster sizes.}
\label{fig:lumi_asym}
\end{figure}
A typical lumi detector asymmetry width is 200 ppm for nominal running conditions. 
Fig. \ref{fig:lumi_asym} shows the variation of lumi 1 asymmetry width with the 
beam raster size.
Six of the lumis, 1-6, used photomultiplier tubes and two, 7 and 8, used 
vacuum photodiodes, although lumi 7 was malfunctioning during the measurements
discussed here. Each lumi detector
was used as an independent measurement, no combinations were made. The lumis
have better statistical precision than the G$^0$ FPD and they have no deadtime (as 
their signals are integrated), which make them more suitable
for measuring target density related effects. However, the lumis are also very 
sensitive to beam related effects that are independent of the target, like 
scraping and halo. The figures that follow are from analysis of the data taken 
during the second engineering run.

A model for target density fluctuations was used to interpret the results 
obtained from data taken at constant beam current, constant pump speed and 
varying the beam raster. The asymmetry in a quartet is computed with
\begin{equation}\label{asym}
A = \frac{N^+ - N^-}{N^+ + N^-} = \frac{\Delta N}{N} 
\end{equation}
where N$^+$ and N$^-$ are the number 
of detector hits per unit beam charge in the corresponding helicity state of a 
quartet. The statistical width of the asymmetry in eq. \ref{asym} is given by 
$\sigma^2_0 = 1/N$.
A source of noise that acts on the time scale of the helicity change will
result in an additional contribution to the asymmetry width resulting in a 
measured width of
\begin{equation}\label{sig}
\sigma_m^2 = \sigma_0^2 + \sigma_b^2
\end{equation}
where $\sigma_0$ is the part of the detector asymmetry width associated with
statistical fluctuations and $\sigma_b$ is that associated with noise. 
In this model the assumption is that $\sigma_b$ is dominated by target density 
fluctuations and depends on the beam raster size as an inverse power law
\begin{equation}\label{sigb}
\sigma_b = \frac{\sigma_{\rho}}{r^x}
\end{equation}
where $\sigma_{\rho}$ is independent of raster size and r$^x$ is the raster linear size
to a power that can be a fit parameter or forced by a model to a specific value. 
Eq. \ref{sigb} was compared to a fit with fixed raster size exponent $x =$ 2, based on
heuristic arguments that the extra width on the measured asymmetry is proportional
to the size of the target density fluctuations and that target density fluctuations 
are inversely proportional to the target volume illuminated by the beam (which is 
the target length times raster area), $\sigma_b \sim \delta \rho \sim$ 1/A$_{raster}$.
Although the heuristic approach may be crude, it is a useful comparison. 
\begin{figure}[htb]
\centering
\rotatebox{-90}
{\includegraphics[scale=0.3]{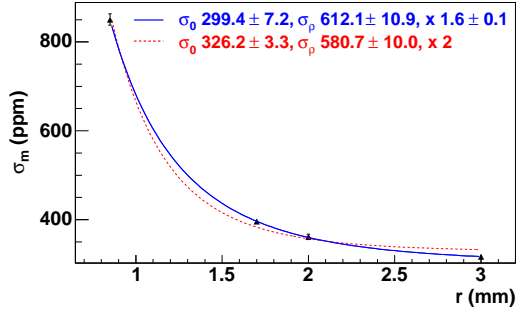}}
\caption{The measured G$^0$ FPD asymmetry width at 40 $\mu$A beam and 31 Hz pump speed.}
\label{fig:4031boil_raw}
\end{figure}
\begin{figure}[htb]
\centering
\rotatebox{-90}
{\includegraphics[scale=0.3]{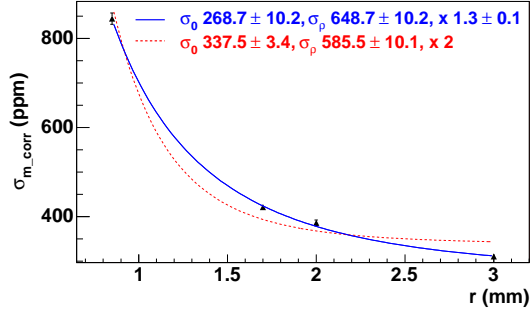}}
\caption{The measured G$^0$ FPD asymmetry width linearly regressed for beam parameters at
40 $\mu$A beam and 31 Hz pump speed.}
\label{fig:4031boil_corr}
\end{figure}

The measured asymmetry width versus raster size (see fig. \ref{fig:4031boil_raw}, 
for example) displays a pronounced nonlinear behavior of the asymmetry width as 
a function of the raster size at constant beam current and constant pump speed. 
Although the exact analytical dependence of the asymmetry width on the raster 
size is unknown, it was found that an inverse power law, eq. \ref{sigb},
fits the data well at 40 and 30 $\mu$A beam current. At lower beam currents the
inverse power law works poorly, while the fixed power law ($x =$ 2) works
better. This may be due to the fact that as the beam intensity reduces, the density
fluctuations decrease and the density becomes more or less constant on the
time scale of a helicity state (30 Hz) over a wide range of raster sizes (which means
that the heuristic arguments outlined above are holding better at lower beam currents).
Both approaches were used to extract the density fluctuations versus raster size and
they were compared. Fig. \ref{fig:4031boil_raw} has the measured G$^0$ FPD 
asymmetry width versus raster size with the beam at 40 $\mu$A and the pump rotating 
at 31 Hz. Fig. \ref{fig:4031boil_corr} has the asymmetry width in the same conditions 
as  fig. \ref{fig:4031boil_raw}, but the asymmetries have been linearly regressed
for beam parameter correlations (like positions and angles at the target, 
charge and energy).
The solid line is the fit with three parameters ($\sigma_0$, $\sigma_{\rho}$, 
and $x$), while the dashed line is the fit for the heuristic model, with the 
raster size exponent fixed at $x =$ 2.
\begin{table}[ht]
\centering
\caption{Density fluctuation results at 40 $\mu$A beam current and 31 Hz
pump speed.}
\vspace{0.3cm}
\begin{tabular}{c c c | c c}
\hline\hline
Detector  &  $x$ & $\sigma_b$ & $x_{corr}$  & $\sigma_{b\_{corr}}$ \\ [0.5ex]
\hline
FPD       &  2      & 145.2   & 2           &  146.4 \\
          & 1.61    & 199.9   & 1.3         & 266.2 \\
\hline
Lumi 1-4  &  2      & 177.8   & 2           & 209.3 \\
          & 1.61    & 236.6   & 1.56        & 295.8 \\
\hline
Lumi 5-8  &  2      & 168.6   & 2           & 173.4 \\
          & 1.55    & 237.9   & 1.48        & 304.4 \\ [1ex]
\hline
\end{tabular}
\label{tab:sum}
\end{table}
The results for the G$^0$ FPD and the lumis at the nominal running point 
are summarized in Table \ref{tab:sum}, where $\sigma_b$ is 
extrapolated with eq. \ref{sigb} to the nominal raster size of 2 mm (asymmetry width 
unit is ppm, fitting errors have been omitted for clarity).

\begin{figure}[htb]
\centering
\rotatebox{-90}
{\includegraphics[scale=0.3]{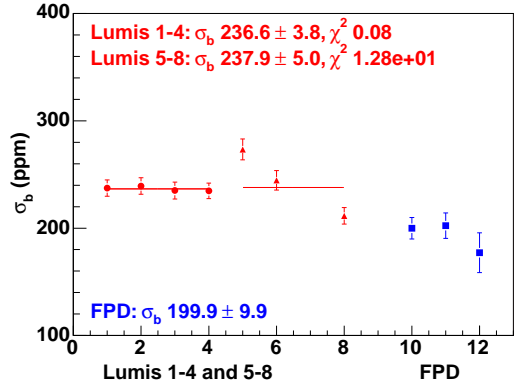}}
\caption{Uncorrected target boiling contribution to asymmetry width 
at nominal running conditions (beam 40 $\mu$A, pump 31 Hz, raster size 2 mm).}
\label{fig:asym_raw}
\end{figure}
\begin{figure}[htb]
\centering
\rotatebox{-90}
{\includegraphics[scale=0.3]{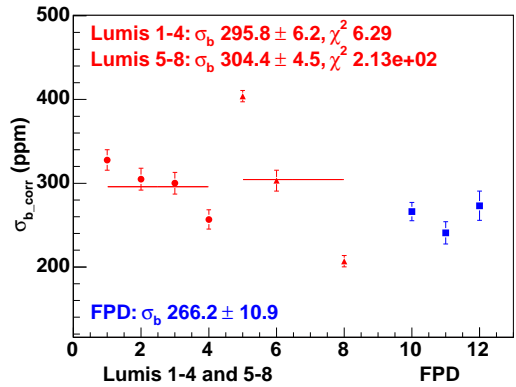}}
\caption{Beam corrected target boiling contribution to asymmetry
width at nominal running conditions.}
\label{fig:asym_corr}
\end{figure}
Figs. \ref{fig:asym_raw} and \ref{fig:asym_corr} are a summary of the 
model extracted, eq. \ref{sigb},
target density fluctuations for the lumi and G$^0$ FPD at the
nominal running point (beam 40 $\mu$A, raster area 2x2 mm$^2$, and pump 31 Hz).
Fig. \ref{fig:asym_raw} shows the extracted target density fluctuations
from the measured detectors asymmetry widths. On the horizontal
axis, numbers 1-8 indicate the respective lumi detector (7 missing), number
10 denotes the G$^0$ FPD, and numbers 11 and 12 denote the segmentation
of the G$^0$ FPD in two independent detectors, to check for concurrence 
of the extracted target effect (the two segments are summed in the analysis as
the G$^0$ FPD). The first 4 lumi detectors are in a remarkable
agreement for the extracted target effect, lumis 5-8 are about 2 sigma
away from each other, but average at the same value as lumis 1-4. The
two segments of the G$^0$ FPD measure similar effects between
themselves (within 1 sigma), and their sum underestimates the $\sigma_b$ 
extracted from the lumis data by 15\%.
Fig. \ref{fig:asym_corr} shows the extracted $\sigma_b$ from the corrected 
detectors asymmetry widths with the linear regression for beam parameters. 
After the linear regression was applied the extracted value for $\sigma_b$  
becomes larger. The increase is about the same, 60-65 ppm, for the average
of lumis 1 to 4, the average of lumis 5 to 8 and the G$^0$ FPD. The
individual data points are much more dispersed compared to the uncorrected
data. Lumi 8 is the only detector whose extracted value for $\sigma_b$
decreases with the regression. In our model, $\sigma_b$, 
defined by eq. \ref{sigb}, encompasses target and beam related effects.
As a consequence it is expected that, after the asymmetry data are regressed
for beam parameters, $\sigma_b$ becomes smaller. In conclusion, it seems
that the linear regression induces noise into the asymmetry measurement
instead of subtracting it at a level of 65 ppm, consistent among the
average of lumi 1-4, lumi 5-8 and the G$^0$ FPD. There are beam
effects that cannot be extracted by a linear regression out of the 
asymmetry data, like scraping and halo. For the nominal running conditions
in the experiment (40 $\mu$A, 3 GeV beam rastered on a square of side 2 mm 
and with the pump rotating at 31 Hz) we quote an upper limit for target 
density fluctuations contribution to the asymmetry width of 238$\pm$65 ppm 
(the highest value in the $\sigma_b$ column in Table \ref{tab:sum}, where the 
uncertainty is taken to be the biggest difference between raw and regressed 
values for $\sigma_b$).
In the normal running mode the asymmetries are measured in 16 Q$^2$
bins. The G$^0$ FPD gets segmented accordingly, and in a typical Q$^2$ bin 
the asymmetry width is about 1200 ppm. A target related effect of 238 ppm 
on top of this nominal width increases the Q$^2$ bin asymmetry width by 2$\%$,
a negligible systematic effect for this experiment.

\begin{figure}[htb]
\centering
\rotatebox{-90}
{\includegraphics[scale=0.3]{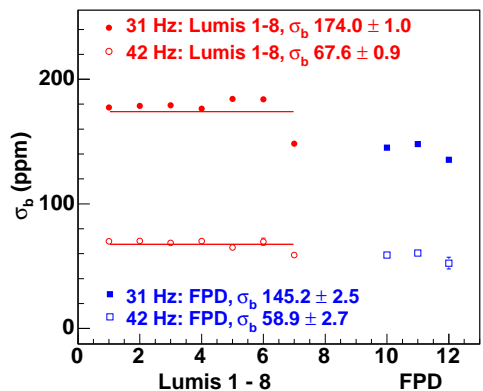}}
\caption{Target boiling contribution to asymmetry width with 
the heuristic approach. Comparison between 42 and 31 Hz pump 
speeds, beam at 40 $\mu$A, raster size 2 mm.}
\label{fig:asym_eurist}
\end{figure}
Fig. \ref{fig:asym_eurist} shows the extracted values for $\sigma_b$ at 40 $\mu$A 
beam and two pump speeds, 31 and 42 Hz. Since there weren't enough
data to make a model extraction (using eq. \ref{sigb}) of the target
density effect, the data for the two pump speeds have been compared to 
the heuristic approach ($x =$ 2). The horizontal axis is the same as in
figs. \ref{fig:asym_raw} and \ref{fig:asym_corr}. The heuristic approach seems to
extract consistent values for $\sigma_b$, at the same pump speed, among
lumis and the two independent segments of the G$^0$ FPD. In this
approach a significant drop in the target density fluctuations is seen
between the two states of the pump, 31 and 42 Hz, suggesting that the
boiling contribution drops with a power of the frequency for this
target (the power law favored by this approach is $\sigma_b \sim 1/f^3$.

The results from the first engineering run were analyzed with the
heuristic approach, as there were not sufficient data at each beam 
current to make a fit with three parameters. However, those results 
are consistent with the results extracted with the same approach 
from the second engineering run. The target cell was replaced 
in between runs. Although the target material was the same for both, 
the exit window of the second cell was reduced by a factor of 3.6, 
from 0.28 mm to 0.08 mm. Within the limits of the model, the density 
fluctuations in a target with high speed longitudinal flow do not seem 
influenced by the exit window heating.

There is one caveat due to a beam property for all the numbers quoted in 
Table \ref{tab:sum}. Over the period of time when these target data were 
taken, the intrinsic beam spot size (unrastered) was 0.107 mm in x and
0.03 mm in y, while typically it is measured to be about ten times bigger
in area. In these conditions the instantaneous power density on the target 
was ten times higher during the target density fluctuations studies than it is
during normal running. The raster spreads it uniformly over the raster area
on a time scale of 25 kHz, while the intrinsic beam repetition rate is 31 MHz.
There are no studies to support the influence of the unrastered beam size
over target density fluctuations.

A study has been done to check the variation of the lumis asymmetry 
widths with the target mean temperature. Changing the target mean 
temperature between 18 and 21~K  resulted in a 3\% change in the lumis 
asymmetry widths [8], which is negligible. This supports the conclusion 
that the density fluctuations mechanism in this target is not dominated 
by micro-bubble formation.

\subsubsection{Density reduction}

\begin{figure}[htb]
\centering
\rotatebox{-90}
{\includegraphics[scale=0.3]{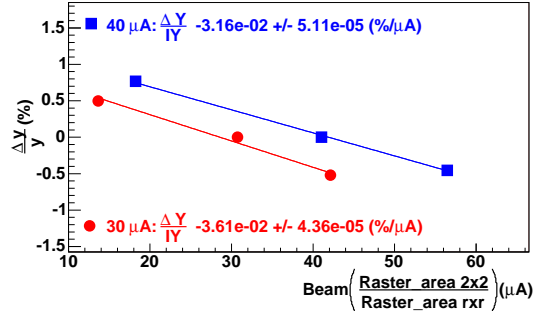}}
\caption{FPD relative normalized yield versus extrapolated beam current.}
\label{fig:yslope}
\end{figure}

To extract the global detector yield reduction versus beam current, and avoid 
deadtime corrections, data at the same beam current and different raster sizes
were extrapolated to nominal raster size, different beam currents, but the same 
power density. In this approach a beam intensity of 40 $\mu$A and raster of 1 mm 
was assumed the same as 160 $\mu$A and raster of 2 mm, since in both situations 
the power density is the same on the target.
The slopes at two different beam currents, 40 and 30 $\mu$A, were measured
versus beam extrapolated currents from different raster sizes and the results are
shown in fig. \ref{fig:yslope} as relative yield change versus beam current.

\begin{figure}[htb]
\centering
\rotatebox{-90}
{\includegraphics[scale=0.3]{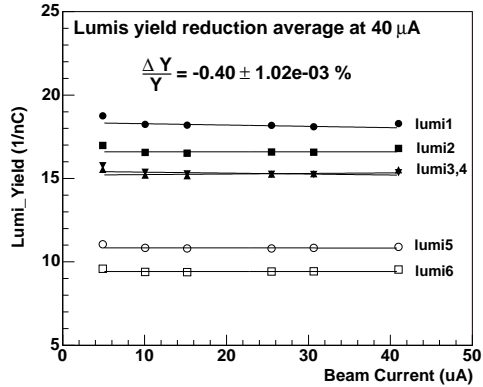}}
\caption{Lumis 1-6 normalized yield versus beam current.}
\label{fig:ylslope}
\end{figure}
At a pump speed of 31 Hz, the FPD yield slopes measured with this indirect method are 
about the same, the 40 $\mu$A measurement yielded -3.16e-2 $\pm$ 5.11e-5 ($\%$/$\mu$A) 
and the 30 $\mu$A measurement yielded -3.61e-2 $\pm$ 4.36e-5 ($\%$/$\mu$A). 
The yield reduction was also measured for the lumi detectors. In this case 
the scan was done directly versus beam current, since these detectors are
deadtime free, and the results are shown in fig.~\ref{fig:ylslope} as the 
relative change in normalized yield as a function of beam current.
From these measurements it can be inferred that the global yield reduction 
when the pump rotates at 31 Hz is less than 1.5$\%$ at 40 $\mu$A beam on 
target for the FPD and 0.4\% for the lumis. The same yield slope extraction 
approach at a pump speed of 42 Hz, 
yielded slopes one order of magnitude smaller than the slopes at 31 Hz. The 
global FPD yield reduction, due to the target, in the nominal running conditions 
is tolerable for this experiment. 

\section{Summary}

In ten months of operational experience in the forward angle mode the target 
has performed reliably,
without a single major incident. At the nominal beam of 40 $\mu$A and 
spot size of 2 mm, with the pump running at 31 Hz, target density fluctuations 
increase the asymmetry width by at most 2\%. In the same nominal conditions 
the global relative yield reduction of the G$^0$ FPD due to the target 
density reduction, has an upper limit of 1.5\%. 
The part of the asymmetry width due to target density fluctuations varies 
nonlinearly with inverse pump speed (constant beam conditions). The target
density fluctuations mechanism is not dominated by micro-bubble formation and
by heating at the exit window. The target has a capacity of at least 1000 W 
power load while staying 3 K subcooled liquid. The effects produced by the 
target satisfy the requirement of negligible systematic effects in a 
parity-violation experiment.

\section{Acknowledgments}

This project was supported by the National Science Foundation under Grant No.
PHY-0244899.
The authors wish to thank the Polarized Target Group at Jefferson Lab for
the outstanding technical support during target testing and installation,
especially drs. Mike Seely and Dave Meekins.



\begin{thebibliography}{00}

\bibitem[1]{g0prop} The G$^0$ Collab.,  G0 Technical Design Report, Jefferson Lab E91-017,
December, 1993.

\bibitem[2]{tgtprop} C.E. Jones and E.J. Beise, The G0 Liquid Hydrogen Target Preliminary Design
Document,G0-collab. Internal Report November 25, 1998.

\bibitem[3]{beise} E.J. Beise et al., Nucl. Instr. and Meth. {\bf A 378}, 383 (1996).

\bibitem[4]{controls} E.J. Beise et al., The G0 Target Controls Manual, revised February 5,2004.

\bibitem[5]{userman} R.Carr and S. Covrig, The G0 Target User's Guide, October 14, 2002.

\bibitem[6]{heattr} See for example, H.Y.Wong, Handbook of essential formulae and data on heat transfer for engineers, London, Longman, 1977.


\bibitem[7]{greg} G. Smith, p{\it rivate comm.}, May 2004.


\end{thebibliography}
\end{document}